\RequirePackage[hyphens]{url}
\documentclass[conference]{IEEEtran}
\IEEEoverridecommandlockouts
\usepackage{cite}
\usepackage{amsmath,amssymb,amsfonts}
\usepackage{graphicx}
\usepackage{xcolor}
\usepackage{textcomp}
\usepackage{hyperref}
\usepackage{cleveref}
\Crefformat{figure}{#2Fig.~#1#3}
\Crefmultiformat{figure}{Figs.~#2#1#3}{ and~#2#1#3}{, #2#1#3}{ and~#2#1#3}
\usepackage{siunitx}  
\usepackage{todonotes}
\usepackage{xspace}
\usepackage{algorithm}
\usepackage{algpseudocode}
\usepackage{caption}
\usepackage{subcaption}
\usepackage{pgfplots}
\usepackage{tabularx}
\pgfplotsset{compat=newest}
\usepackage{tikz}
\usepackage[right]{eurosym}
\usepackage{flushend}
\usepackage[normalem]{ulem}

\usepackage[nolist]{acronym} %

\usepackage{ifthen}
\newboolean{usetodonotes}
\setboolean{usetodonotes}{false}

\newboolean{cameraready}
\setboolean{cameraready}{true}

\newboolean{arxiv}
\setboolean{arxiv}{true}

\usepackage{eso-pic}
\usepackage{setspace}
\ifthenelse{\boolean{arxiv}}{
\AddToShipoutPicture{%
	\ifnum\value{page}=1
	\AtPageLowerLeft{%
		\raisebox{2.5\baselineskip}{\makebox[\textwidth][l]{\hspace{1.78cm}\begin{minipage}{8.2cm}
					\begin{spacing}{0.75}
						{\footnotesize\textcolor{black!60}{\textcopyright~2022 IEEE. Personal use of this material is permitted. Permission from IEEE must be obtained for all other uses, in any current or future media, including reprinting/republishing this material for advertising or promotional purposes, creating new collective works, for resale or redistribution to servers or lists, or reuse of any copyrighted component of this work in other works.}}
					\end{spacing}
		\end{minipage}}}%
	}%
	\AtPageLowerLeft{%
		\raisebox{3.5\baselineskip}{\makebox[\textwidth][l]{\hspace{11.6cm}\begin{minipage}{8.2cm}
					\begin{spacing}{0.75}
						{\footnotesize\textcolor{black!60}{\textit{Note:} This is the authors' version of the article accepted for publication at IEEE International Conference
								on Physical Assurance and Inspection of Electronics (PAINE 2022).}}
					\end{spacing}
		\end{minipage}}}%
	}%
	\else
	\AtPageLowerLeft{%
		\raisebox{3\baselineskip}{\makebox[\textwidth][l]{\hspace{1.8cm}\begin{minipage}{8.8cm}
					\begin{spacing}{0.75	}
						{\footnotesize\textcolor{black!60}{\textcopyright~2022 IEEE}}
					\end{spacing}
		\end{minipage}}}%
	}%
	\fi
}
}{}

\newcommand\addplotgraphicsnatural[2][]{
    \begingroup
    \pgfqkeys{/pgfplots/plot graphics}{#1}
    \setbox0=\hbox{\includegraphics{#2}}
    \pgfmathparse{\wd0/(\pgfkeysvalueof{/pgfplots/plot graphics/xmax} - \pgfkeysvalueof{/pgfplots/plot graphics/xmin})}
    \let\xunit=\pgfmathresult
    \pgfmathparse{\ht0/(\pgfkeysvalueof{/pgfplots/plot graphics/ymax} - \pgfkeysvalueof{/pgfplots/plot graphics/ymin})}
    \let\yunit=\pgfmathresult
    \xdef\marshal{%
        \noexpand\pgfplotsset{unit vector ratio={\xunit\space \yunit}}%
    }
    \endgroup
    \marshal
    \addplot graphics[#1] {#2};
}  

\ifthenelse{\boolean{usetodonotes}}{
	\paperwidth=\dimexpr \paperwidth + 6cm\relax
	\oddsidemargin=\dimexpr\oddsidemargin + 3cm\relax
	\evensidemargin=\dimexpr\evensidemargin + 3cm\relax
	\marginparwidth=\dimexpr \marginparwidth + 3cm\relax
	\presetkeys
	{todonotes}
	{
		size=\footnotesize,
		caption={},
		color=red!60
	}{}
}{
	\presetkeys
	{todonotes}
	{disable}{}
}

\newcommand{\raspi}{Raspberry Pi\xspace}

\def\BibTeX{{\rm B\kern-.05em{\sc i\kern-.025em b}\kern-.08em
    T\kern-.1667em\lower.7ex\hbox{E}\kern-.125emX}}
\begin{document}

\title{EM-Fault It Yourself: Building a Replicable EMFI Setup for Desktop and Server Hardware}

\author{
\IEEEauthorblockN{Niclas Kühnapfel\IEEEauthorrefmark{1},
	Robert Buhren\IEEEauthorrefmark{1},
	Hans Niklas Jacob\IEEEauthorrefmark{1},
	Thilo Krachenfels\IEEEauthorrefmark{1},\\
	Christian Werling\IEEEauthorrefmark{1},
	Jean-Pierre Seifert\IEEEauthorrefmark{1}\IEEEauthorrefmark{2}}
	\IEEEauthorblockA{\IEEEauthorrefmark{1} Technische Universit\"at Berlin, Chair of Security in Telecommunications, Germany\\
	}
	\IEEEauthorblockA{\IEEEauthorrefmark{2} Fraunhofer SIT, Germany\\
	}
}

\maketitle

\ifthenelse{\boolean{cameraready}}{
	\ifthenelse{\boolean{arxiv}}{
		\thispagestyle{plain}
		\pagestyle{plain}
	}{}
}{
	\thispagestyle{plain}
	\pagestyle{plain}
}

\begin{acronym}
    \acro{EM}[EM]{electromagnetic}
    \acro{EMFI}[EMFI]{electromagnetic fault injection}
    \acro{FI}[FI]{fault injection}
    \acro{ROT}[RoT]{root of trust}
    \acro{DUT}[DUT]{device under test}
    \acro{SOC}[SoC]{system on a chip}
    \acroplural{SOC}[SoCs]{systems on a chip}
    \acro{PCB}[PCB]{printed circuit board}
    \acro{IC}[IC]{integrated circuit}
    \acro{AMD-SP}[AMD-SP]{AMD Secure Processor}
    \acro{STIM}[STIM]{solder thermal interface material}
    \acro{IP}[IP]{intellectual property}
    \acro{DRM}[DRM]{digital rights management}
    \acro{ARK}[ARK]{AMD root key}
    \acro{FPGA}[FPGA]{field-programmable gate array}
\end{acronym}

\acresetall

\begin{abstract}
\Ac{EMFI} has become a popular \ac{FI} technique due to its ability to inject faults precisely considering timing and location.
Recently, ARM, RISC-V, and even x86 processing units in different packages were shown to be vulnerable to \ac{EMFI} attacks.
However, past publications lack a detailed description of the entire attack setup, hindering researchers and companies from easily replicating the presented attacks on their devices.

In this work, we first show how to build an automated \ac{EMFI} setup with high scanning resolution and good repeatability that is large enough to attack modern desktop and server CPUs.
We structurally lay out all details on mechanics, hardware, and software along with this paper.
Second, we use our setup to attack a deeply embedded security co-processor in modern AMD \acp{SOC}, the \ac{AMD-SP}.
Using a previously published code execution exploit, we run two custom payloads on the AMD-SP that utilize the \ac{SOC} to different degrees.
We then visualize these fault locations on \ac{SOC} photographs allowing us to reason about the \ac{SOC}'s components under attack. 
Finally, we show that the signature verification process of one of the first executed firmware parts is susceptible to \ac{EMFI} attacks, undermining the security architecture of the entire \ac{SOC}.
To the best of our knowledge, this is the first reported \ac{EMFI} attack against an AMD desktop CPU.
\end{abstract}


\acresetall
\section{Introduction}
Nowadays, hardware security is an important requirement for chip vendors because all kinds of processors and \acp{SOC} are commonly used as trust anchor for devices and platforms.
Immutable and sometimes even secret code or data is embedded into chips to verify and authenticate the boot process, protect \ac{IP}, and enable technologies like \ac{DRM} or trusted computing.
Modern \acp{SOC} include dedicated processor cores to handle security-related tasks and to protect digital secrets from the main cores, which execute untrusted and potentially vulnerable software components.
Intel's Converged Security and Management Engine (CSME) and AMD's Secure Processor (AMD-SP) are security processors integrated into the main CPU die \cite{amd_psp_whitepaper, intel_csme_whitepaper}.
Their attack surface is reduced by restricted communication interfaces, separated resources, and authenticity and integrity checks of code and data through cryptographic signatures.
Nevertheless, recent research has shown that these dedicated security processors can be vulnerable to hardware attacks, such as \ac{FI}~\cite{buhren_one_2021}. 

\Ac{FI} attacks exploit the essential operating conditions of electronics required for their intended functionality.
Altering the supply voltage or clock frequency and inducing energy by optical or \ac{EM} effects may destabilize an integrated circuit or processor so that faults occur.
As  a prominent and easy-to-use technique, voltage glitching was applied to, e.g., circumvent firmware protection~\cite{bozzato_shaping_2019, buhren_one_2021, bittner_forgotten_2021} and attack secure enclaves~\cite{chen_voltpillager_2021, kenjar_v0ltpwn_2020, qiu_voltjockey_2020} on modern \acp{SOC} and CPUs. 
While voltage glitching affects the entire \ac{DUT}, laser and \ac{EMFI} are more precise and only affect a specific area of the \ac{DUT} by locally injecting charge carriers or applying a quickly changing electromagnetic field, respectively.
Furthermore, both techniques do not require an electrical connection to the power delivery network, potentially allowing non-invasive attacks.
While laser \ac{FI} requires optical access to the decapsulated chip backside, \ac{EMFI} can be leveraged through a non-shielding package.
When attacking a processor, both laser \ac{FI} and \ac{EMFI} are powerful methods to skip or alter instructions and change the software execution flow~\cite{smartphone_lfi, Dutertre2019ExperimentalAO, trouchkine_electromagnetic_2021}.

Traditional \ac{EMFI} attacks exhibit three problems: Firstly, they require more complex setups than, e.g., voltage glitching, making the reliable reproduction of research results more challenging. Such setups, secondly, often comprise proprietary and/or undocumented components, hindering the reproduction of specific attacks and the reuse of hardware components for related use cases, such as laser \ac{FI}. Thirdly, they often lack a large working area for attacking desktop and server hardware.

A typical setup consists of a pulse generator, an XYZ table, a motion controller, and software to orchestrate all components.
Apart from self-built EM pulse generators~\cite{design_val_emfi_platform, abdellatif_silicontoaster_2020, badfet, delarea_lowcost_emfi}, commercial devices are used frequently and were characterized and compared in~\cite{Toulemont2020ASP}.
For repeatability reasons, a setup additionally provides means of calibrating the probe position.
However, such a sophisticated EM setup built from commercial off-the-shelf hardware has, to the best of our knowledge, not been described in past publications.
On the other hand, commercially available setups do not support larger targets, such as server and desktop motherboards altogether (Riscure EM Probe Station 5 \cite{riscure_probe_station}) and lack the possibility to modify most of their parts.

\textbf{Our Contributions.}
Consequently, this work contains the following two main contributions.
Firstly, we build a setup for \ac{EMFI} incorporating the NewAE ChipShouter as a pulse generator, a high-resolution stage large enough to target desktop and server motherboards, a motion controller that drives the stepper motors, and custom control software running on the main control unit.
Moreover, the setup includes positioning cameras for calibration. 
We describe in detail how our setup was built, which components were installed and how the setup works.
Furthermore, we publish all necessary CAD and design files as well as the source code of the control software in order to replicate and build upon our setup.

Secondly, we use our setup to target the \ac{AMD-SP} inside a second-generation Zen AMD Desktop CPU (Zen+) and demonstrate which areas of the processor are susceptible to \ac{EMFI} and what types of faults can be injected.
Since EM radiation is inherently shielded by conductive parts of the outer processor assembly, we show how to prepare the device for \ac{EMFI}.

The conducted experiments demonstrate that the intended execution flow of code running on the ARM-based subsystem can be altered to break out of a loop or even to bypass critical checks.
Data stored in SRAM is susceptible to \ac{EMFI} as well.
Our target had previously been attacked using voltage \ac{FI} to bypass a security-critical check in the boot loader, which gives an adversary full control over the device~\cite{buhren_one_2021}.
We reproduce this attack and demonstrate that \ac{EMFI} offers significantly higher success rates than voltage \ac{FI}.
To the best of our knowledge, this is the first \ac{EMFI} against an AMD desktop CPU and the first \ac{EMFI} attack on a subsystem of a modern x86 CPU.
\section{Background}
\label{sec:background}
\begin{figure}[tb]
    \centering
    \input{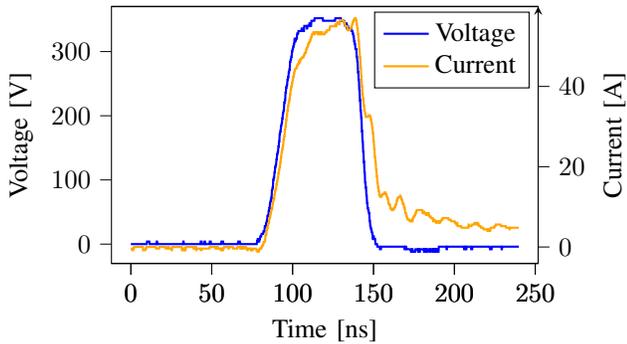}
    \caption{Voltage and current of a pulse generated by the ChipShouter with a \SI{4}{\milli\meter} injection probe tip}
    \label{fig:em_pulse}
\end{figure}

\subsection{Electromagnetic Fault Injection (EMFI)}
Electromagnetic fault injection leverages quickly changing \ac{EM} fields to trigger faults in a device in close proximity. 
As such, it belongs to the class of non-invasive \ac{FI} techniques \cite{cryptoeprint:2005/388}.
This \ac{EM} field induces a voltage into the chip's inner circuits, potentially resulting in transient or persistent faults, e.g., a bit-flip in the SRAM included in an embedded processor.
Most \ac{EMFI} tools follow the same working principle to generate the electromagnetic field \cite{design_val_emfi_platform}:
After a capacitor is charged to a high voltage of up to \SI{1200}{\volt} \cite{abdellatif_silicontoaster_2020}, it is quickly discharged through an injection probe tip.
The high current flowing through the coil of the injection probe tip creates an \ac{EM} field around it and is characterized by its peak voltage, peak current, rise time, fall time, and pulse width as depicted in \Cref{fig:em_pulse}.

Injection probe tips consist of a copper coil with a ferrite core to concentrate the magnetic flux. 
The resulting pulse and \ac{EM} field are determined by the number of windings, winding sense, copper wire diameter, and ferrite diameter \cite{gaine:cea-03657852}.
Therefore, probe tips need to be carefully selected to achieve desired pulse characteristics for a specific target.
Especially, the ferrite diameter and distance to the \ac{DUT}, as well as the number of windings, should be tuned for best results.
Furthermore, targets may require removing the protective cover or heat spreader to minimize the physical gap between the probe tip and the \ac{DUT} \cite{beckers_atmega_emfi}.
However, this process of \emph{decapsulation} or \emph{delidding} turns the theoretically non-invasive \ac{EMFI} attack into a semi-invasive hardware attack \cite{cryptoeprint:2005/388}.

\subsection{AMD Secure Processor}
\label{sub:background:asp}
Since 2013, AMD has embedded the \ac{AMD-SP} into its x86 processors and graphics cards.
It has an ARMv7 core with segregated SRAM and uses an SPI flash as non-volatile storage.
The AMD-SP's proprietary firmware is cohosted on the SPI flash along with the BIOS/UEFI firmware~\cite{amd_psp_whitepaper, buhren_blackhat_2020}

The \ac{AMD-SP} operates before the x86 cores are enabled and boots as follows:
At first, the immutable on-chip bootloader is loaded and executed from the internal ROM.
It then loads the \ac{ARK} from the SPI flash and verifies it by comparing it to an immutable hash.
Afterward, the off-chip bootloader (responsible for the rest of the boot process) is loaded, its \ac{ARK} signature verified, and executed \cite{buhren_one_2021}.
This process aims to ensure the authenticity and integrity of all code run on the \ac{AMD-SP}.

Thus, acting as the root of trust for the whole x86 system, the AMD-SP poses an attractive target for fault injections, effectively compromising the system's chain of trust.
\section{EMFI Setup}
\label{sec:emfi_setup}

\begin{figure}[tb]
    \centering
    \includegraphics[width=0.38\textwidth]{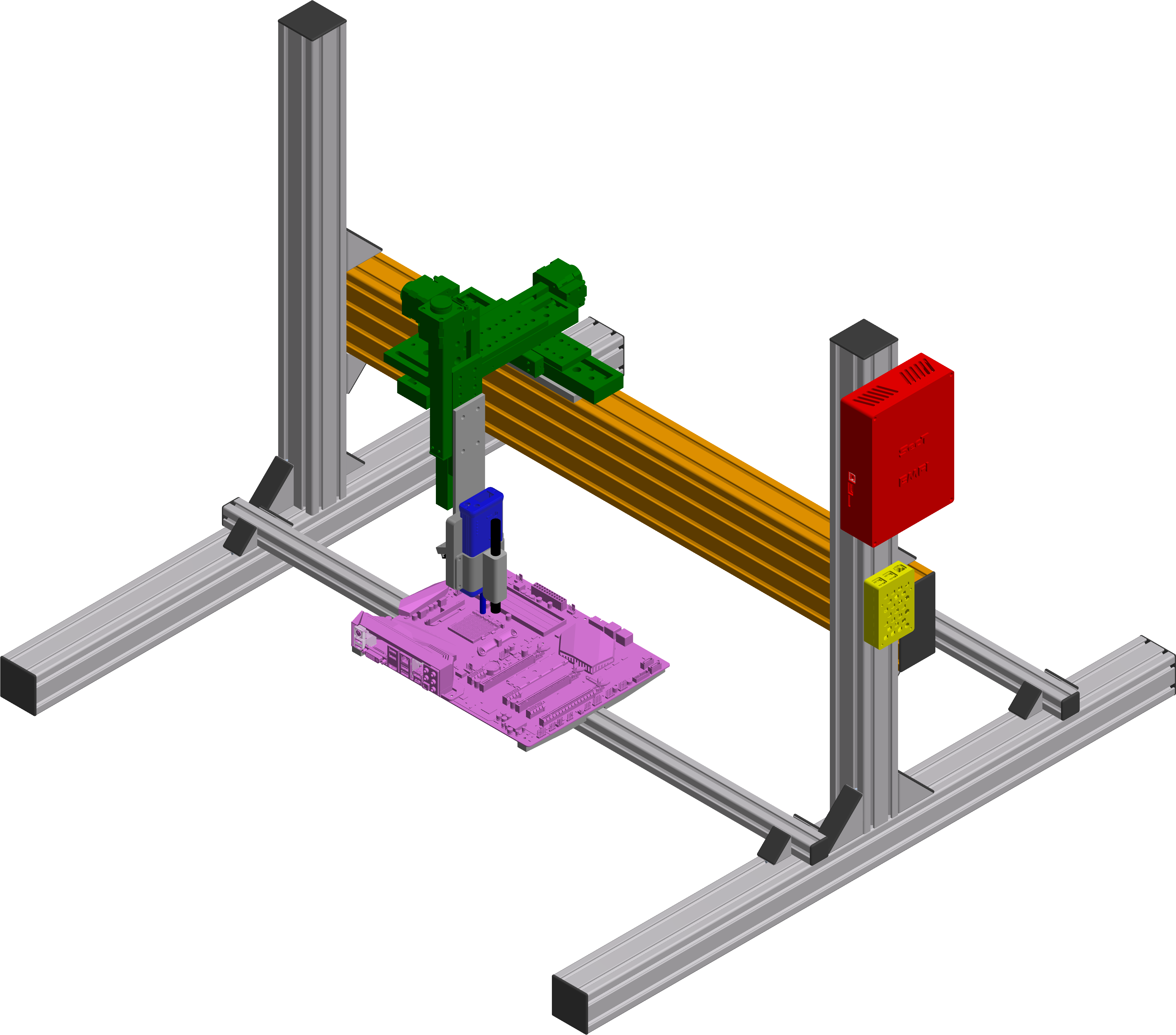}
    \caption{EMFI Setup: XYZ stage (green), motion control unit (red), main control unit (yellow), ChipShouter (blue), target (violet), aluminum crossbar (orange)}
    \label{fig:setup}
\end{figure}

This section describes the general experimental setup built to perform \ac{EMFI} attacks.
All details and source files needed to replicate our setup can be found in the accompanying repository of this work\footnote{\url{https://github.com/fgsect/EM-Fault-It-Yourself}}.
The setup (\Cref{fig:setup}) consists of five main parts: an EM pulse generator (blue), an XYZ precision positioning stage with additional sensors (green), a motion control unit (red), the main control unit (yellow), and the target (violet).

We use NewAE's ChipShouter as an EM pulse generator. 
It can generate high voltage pulses up to \SI{500}{\volt} with a pulse width between \SI{80}{\nano\second} and \SI{960}{\nano\second} \cite{chipshouter_manual}.
Pulses can be triggered by internal trigger logic, an internal programmable pattern generator, or a hardware trigger driven by external circuitry. 
Our setup uses the latter to avoid the inherent delay of software-based trigger logic.
The ChipShouter communicates via a serial interface that allows us to configure all its settings from the main control unit.

Faults injected by electromagnetic fields are location-dependent.
Hence, the XYZ stage is used to move the injection probe tip to different positions. 
The process of probing the \ac{DUT}'s surface at different positions for \ac{EMFI} sensitive areas along a grid is called \emph{scanning} in this work.

Our setup is based on a frame made of aluminum construction profiles that are easily adjustable and extendable.
Three motorized linear stages (Standa 8MT175-100) are placed on the aluminum crossbar (orange), forming an XYZ positioning system.
Each stage has a travel range of \SI{100}{\milli\metre}, an accuracy of \SI{2.5}{\micro\metre} and a maximum speed of \SI{10}{\milli\metre\per\second}.
All three stages combined are remountable along the crossbar, allowing us to place a complete computer and even a rack-mountable server under the crossbar.
We can then move the XYZ stage along the X-axis until its travel range covers a chip's surface.
Smaller targets like computer mainboards or embedded boards can be screwed to the construction profiles using our custom 3D-printed mounting brackets and T-slot nuts.
A 3D-printer board (Bigtreetech SKR Pro v1.2) and its stepper motor drivers (based on Trinamic TMC2209) control the acceleration, speed, and holding torque of each stepper motor.
The board acts as the motion controller, runs a slightly modified version of the open-source 3D-printer firmware Marlin \cite{marlin} and can communicate via a USB-to-UART connection to the main control unit.
As Marlin handles low-level real-time movement and positioning, the main control unit only has to issue high-level commands to move to an absolute or relative position.

\begin{figure}[b]
     \centering
     \begin{subfigure}[t]{0.15\textwidth}
         \centering
         \includegraphics[width=\textwidth]{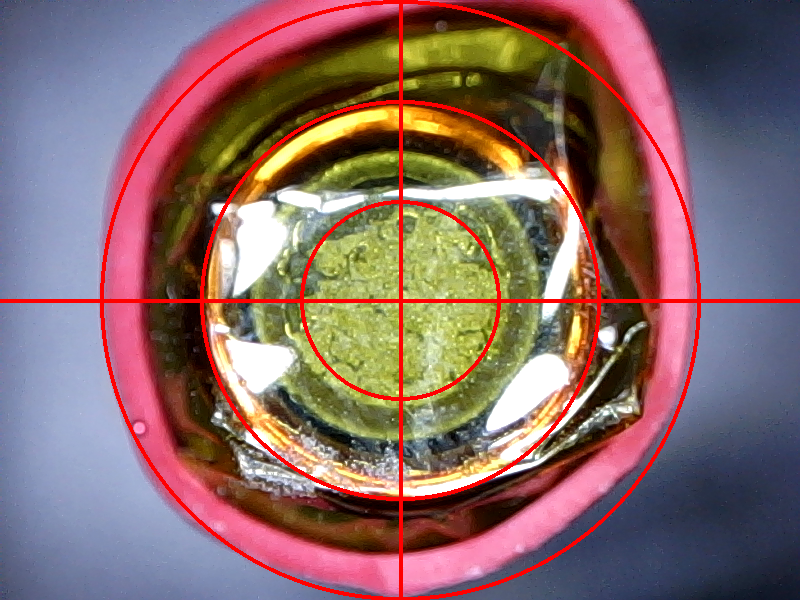}
         \caption{Determining the position of the injection probe tip (calibration camera view)}
         \label{fig:align:injection_probe_tip}
     \end{subfigure}
     \hfill
     \begin{subfigure}[t]{0.15\textwidth}
         \centering
         \includegraphics[width=\textwidth]{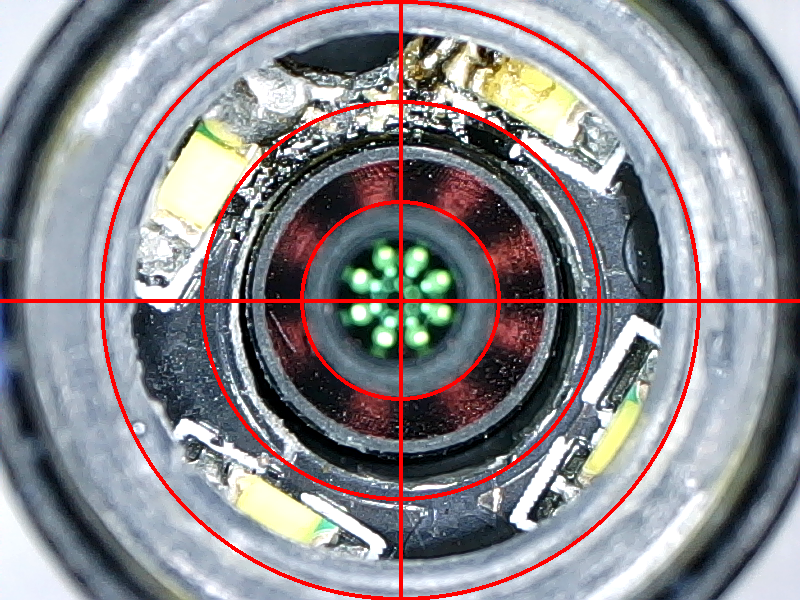}
         \caption{Determining the position of the positioning camera (calibration camera view)}
         \label{fig:align:microscope_cam}
     \end{subfigure}
     \hfill
     \begin{subfigure}[t]{0.15\textwidth}
         \centering
         \includegraphics[width=\textwidth]{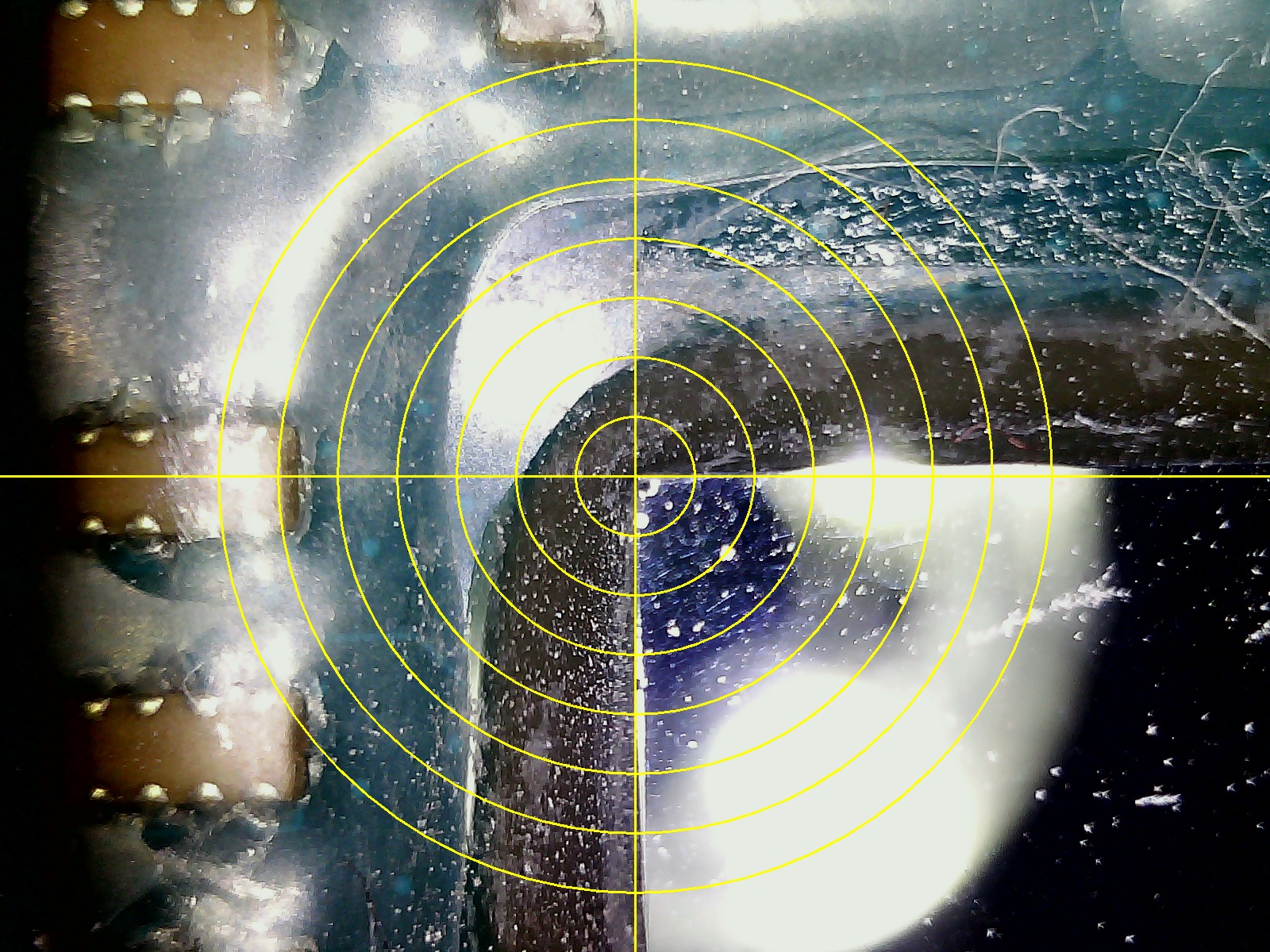}
         \caption{Determining the position of CPU die corner (positioning camera view)}
         \label{fig:align:cpu_die_corner}
     \end{subfigure}
        \caption{Aligning the center of the injection probe tip and the CPU die corner}
        \label{fig:align}
\end{figure}

The pulse generator, a microscope camera, and a thermal camera are mounted on the Z-stage using an aluminum adapter plate and 3D-printed mounting brackets.
The microscope camera (positioning camera) is mounted along the ChipShouter and directed to the bottom to determine the target's position relative to the origin of all stages.
Its position relative to the different injection probe tips has to be measured after every tip change since the probe tips all have a different shape caused by manufacturing tolerances.
To determine the offset between the center of the positioning camera and the center of the injection probe tip, an additional microscope camera (calibration camera) is used.
The calibration camera is temporarily placed below the ChipShouter mount, pointing upwards so that it can provide a view from below to locate the center points of both positioning camera (\Cref{fig:align:microscope_cam}) and injection probe tip (\Cref{fig:align:injection_probe_tip}).
A CPU die corner can then be accurately located using the positioning camera (\Cref{fig:align:cpu_die_corner}).
To align a CPU die corner and the center of the injection probe tip, the exact position of the probe tip can be computed using the offset and the precise position of the die corner.

The maximum Z-depth can be measured by manually moving down in small steps until a piece of paper is nearly jammed between the injection probe tip and \ac{DUT}.
If the attack target needs active cooling, one or multiple fans can be attached to the frame and connected to the motion control unit for speed regulation.
Temperature monitoring is provided by the thermal camera, which is directed to the \ac{DUT}.

All mentioned devices are connected to the main control unit, a \raspi that manages and monitors the devices and orchestrates all actions such as manual movements or attacks.
It runs a custom \ac{EMFI} server application that provides a web interface to control the XYZ stage, start task scripts, as well as stream the positioning, calibration, and thermal camera information.
Attacks and other repetitive tasks are implemented through Python task scripts, making the experiments easily adaptable and reproducible.
They use an internal programming interface provided by the \ac{EMFI} server that allows access to the pulse generator, sensor data, and movement control.

\ac{EMFI} attacks often have a huge parameter search space composed of spatial position, timing, EM pulse characteristics, and possible target-specific parameters.
Exploring these parameters is necessary prior to a successful attack but can take days to weeks.
Hence, it is crucial that our server application tracks and logs all actions and responses of a target.
Moreover, it makes all data available to the scripting API for evaluation.
\section{Attack Setup}
In the following section, we describe how we extended the setup to target the \ac{AMD-SP} of an AMD Ryzen 5 2600 CPU and how the target itself had to be prepared.
These modifications are target-specific and have to be implemented differently for other targets.

The targeted CPU is mounted on an ASUS PRIME X370-PRO mainboard and powered by a standard PC power supply unit.
Several tools are required to replace the co-processor's firmware or to inject commands to its SVI2 bus, which allows voltage control of the x86 cores and the \ac{SOC} \cite{buhren_one_2021}.
We also use a logic analyzer as it can greatly help to debug wiring or signal faults of the SVI2 or SPI bus.

\begin{figure}[tb]
     \centering
     \begin{subfigure}[t]{0.15\textwidth}
         \centering
         \includegraphics[width=\textwidth]{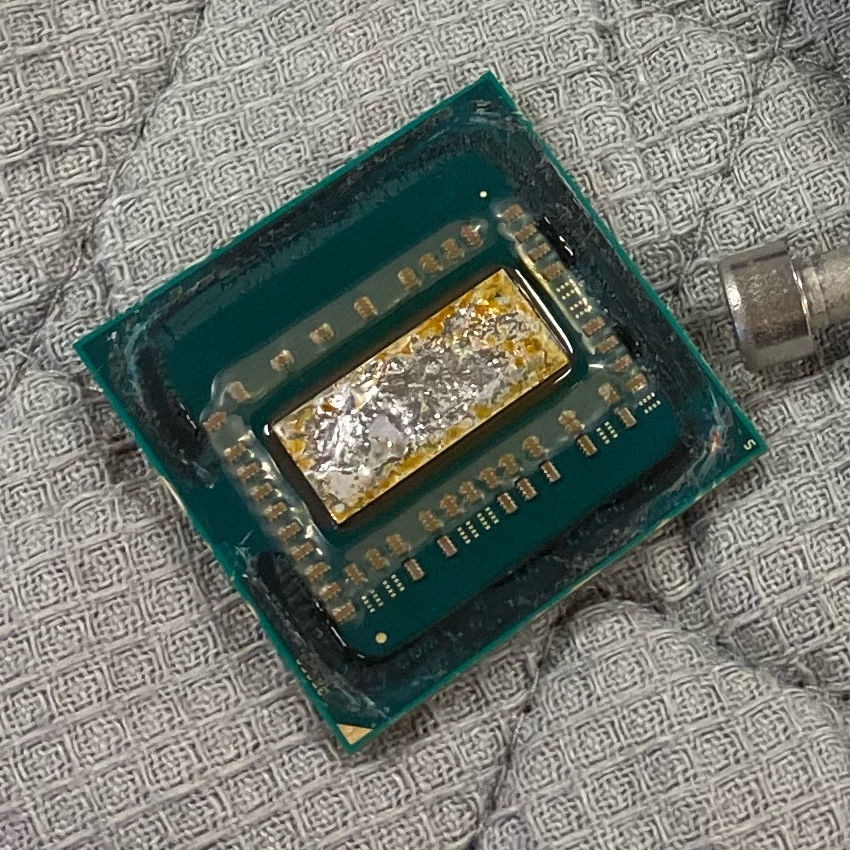}
         \caption{CPU after removing the heat spreader (delidded)}
         \label{fig:delidded_cpu}
     \end{subfigure}
     \hfill
     \begin{subfigure}[t]{0.15\textwidth}
         \centering
         \includegraphics[width=\textwidth]{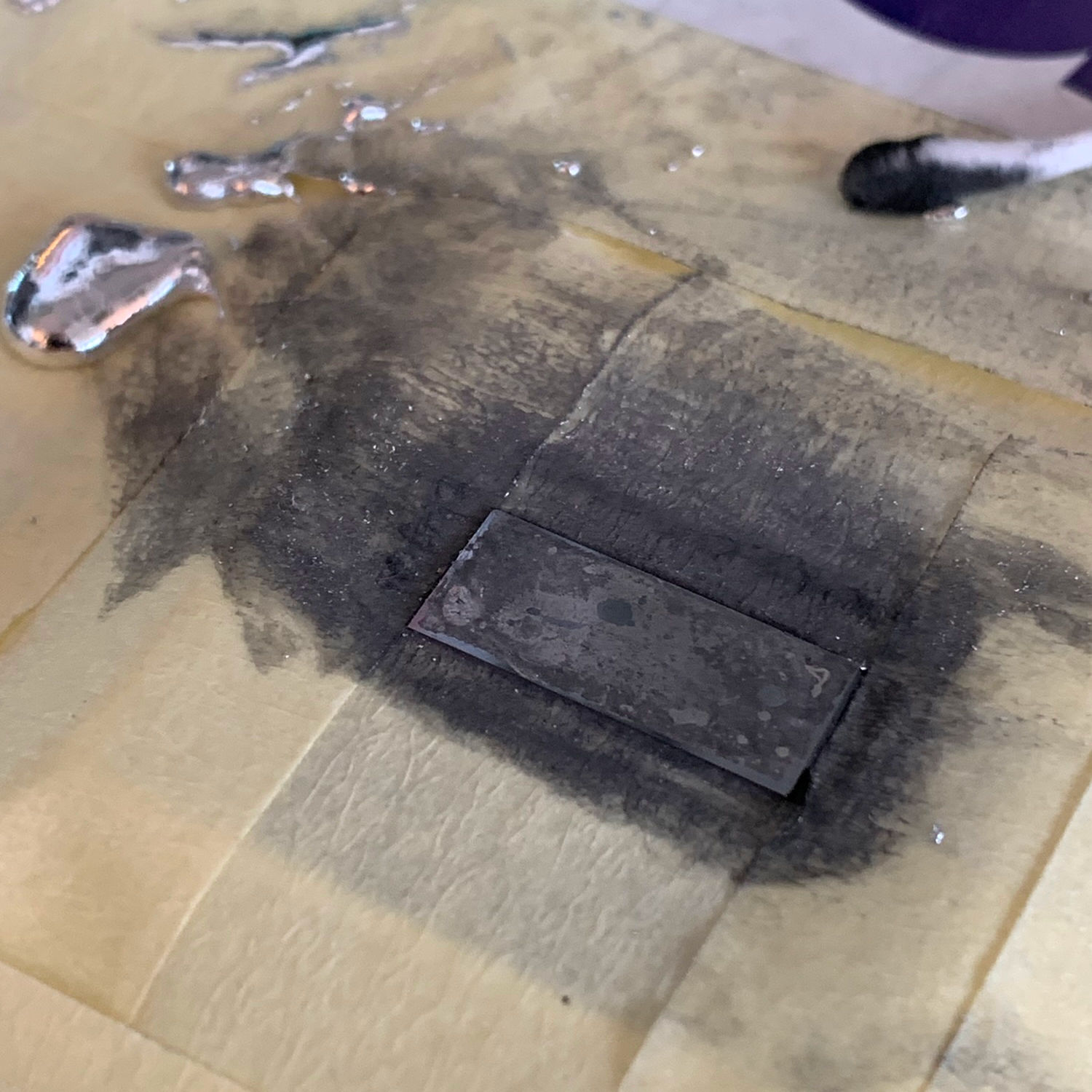}
         \caption{CPU while removing the indium \ac{STIM}}
         \label{fig:tim_removal_cpu}
     \end{subfigure}
     \hfill
     \begin{subfigure}[t]{0.15\textwidth}
         \centering
         \includegraphics[width=\textwidth]{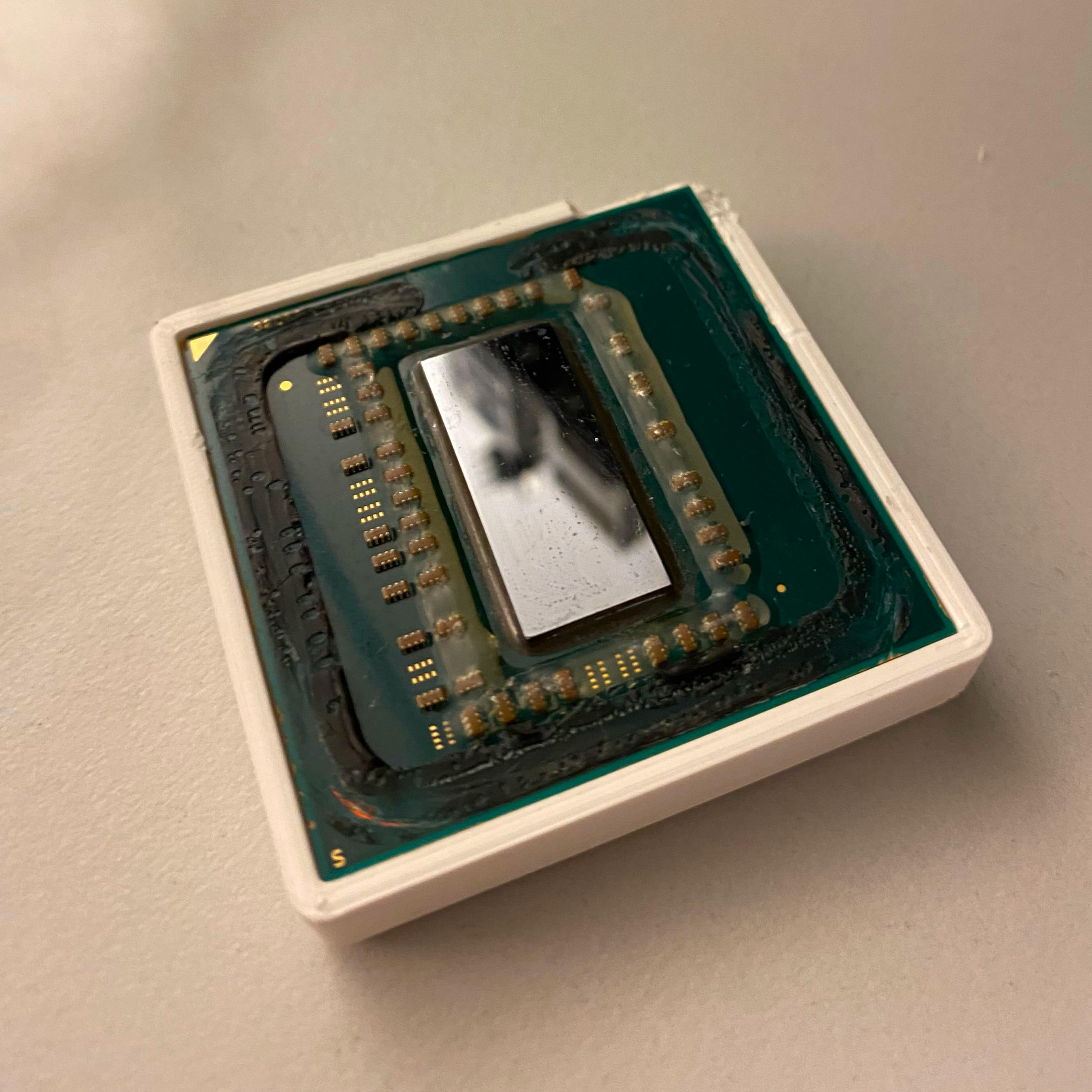}
         \caption{Delidded and cleaned CPU}
         \label{fig:delidded_cleaned_cpu}
     \end{subfigure}
        \caption{Delidding and cleaning of the AMD Ryzen 5 2600 (Zen+)}
\end{figure}

\subsection{Preparation}
The AMD Ryzen 5 2600 is equipped with a glued, indium-soldered nickel-plated copper heat spreader.
This heat spreader shields the \ac{EM} pulses and has to be disassembled, which is called \emph{delidding} or \emph{decapsulation}, and consists of three steps:
First, the glue at the heat spreader's border must be cut.
Second, the processor and heat spreader have to be heated up to a temperature of \SI{200}{\celsius}.
When the indium solder becomes fluid, the processor and heat spreader can be separated using pliers or special tools (e.g., Delid Die Mate 2) in the third step.
Afterward, the die is still covered with the \ac{STIM}, a conductive indium solder shielding the processor die and the \ac{AMD-SP}. 
As depicted in \Cref{fig:tim_removal_cpu}, we used Quicksilver Solder Removal to clean the processor die completely.

\subsection{Supply Voltage}
Previous research showed that lowering an ICs supply voltage can aid \ac{EMFI} attacks \cite{liao_methodology_2019}.
Therefore, we make use of AMD's SVI2 bus to fully control the CPU's V\textsubscript{Core} and V\textsubscript{SoC} voltages \cite{buhren_one_2021}.
Injecting packets onto this bus lets us issue commands to the mainboard's voltage regulators, which drive and monitor these voltages.
Due to its use in prior research \cite{buhren_one_2021}, we used a Teensy 4.0 microcontroller connected to the main control unit for this and other tasks.

\subsection{Trigger}
The correct time to trigger an \ac{FI} attack is often found by searching a limited time frame in relation to some hardware event.
For example, data bus messages or a switching GPIO can serve as a starting point for a trigger.
An additional device is needed to detect and evaluate these events and to delay and forward a trigger signal to the external trigger input of the pulse generator.
While an FPGA would be predestined for this purpose, due to its capability to forward signals instantly, we use the Teensy microcontroller to avoid adding another device to the setup.
The main control unit configures the trigger logic implemented on the Teensy via its USB-to-UART interface.
It monitors the mainboard's SPI bus to precisely time the fault injection and trigger an EM pulse through the hardware trigger input of the pulse generator.

\subsection{Software}
It is not trivial to detect whether faults were injected into a processor running proprietary software, as its behavior is only observable from the outside.
To tackle this problem, we used a vulnerability in the on-chip bootloader to run arbitrary code on the AMD-SP \cite{buhren_blackhat_2020}.
Running our own test code allows us to examine the target for different fault types as described in \Cref{sec:experiments} as well as to use the mainboard's UART and SPI interfaces for communication.

The AMD-SP loads its firmware from the mainboard's SPI flash chip, which also stores the BIOS/UEFI files.
We use a DediProg EM100Pro-G2 flash emulator to replace the soldered flash chip and allow easy and fast modification of the SPI flash contents.
Instead of replacing the SPI flash, it could also be reprogrammed non-invasively using a flash programmer and test clip.
PSPTool \cite{werling_psptool} is able to analyze and modify those flash images easily.

\subsection{Cycle Time}
The cycle time is the amount of time it takes to move to a position, power up the system, run the attack, and fully power down the system so the procedure can start from the beginning.
In some cases, a simple reset is sufficient, and a complete power cycle is not needed.
As some attack parameters may have to be brute-forced, the cycle time should be kept as low as possible.
The fastest way of resetting our target is to hijack the PS\_ON line of the PSU by controlling it through the Teensy microcontroller board.
To power up the target, only PS\_ON of the PSU and PWR\_SW (power switch) of the mainboard have to be enabled successively by the Teensy.
On the other hand, the Teensy only has to disable the PS\_ON pin for a power-down since it completely cuts the power.
\section{Experiments}
\label{sec:experiments}
In this section, we describe the conducted experiments, how we found areas susceptible to our EM pulses, and which faults were successfully injected while running our test code.

\begin{figure*}[t]
    \centering
    \input{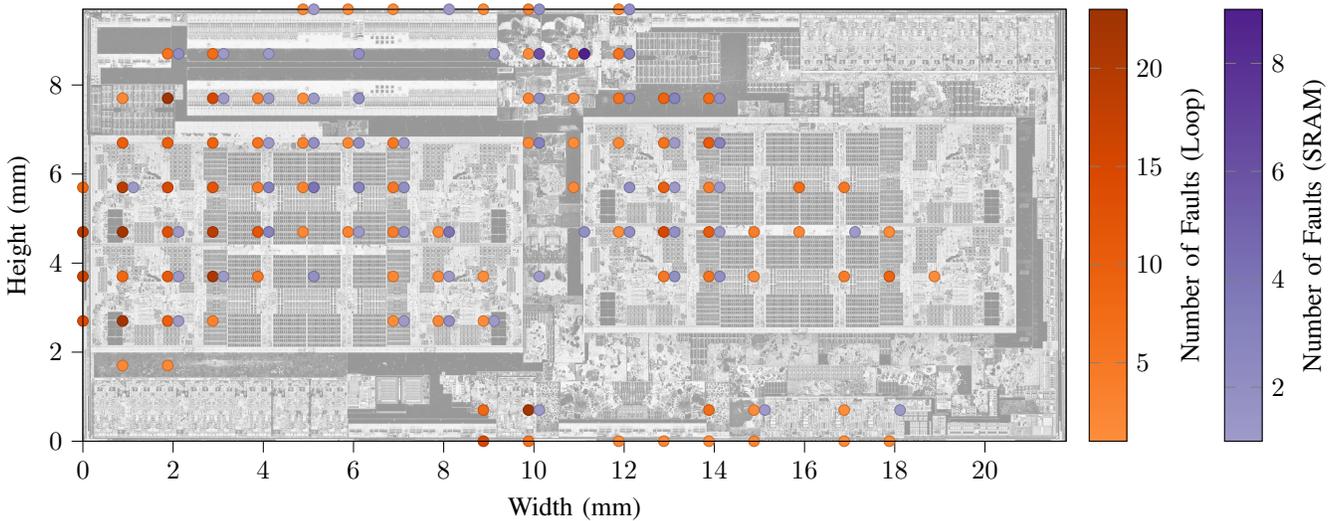}
    \caption{Scan of the \acp{SOC} surface while executing loop and SRAM payload (4mm clockwise injection probe tip; 1mm grid; 100 attempts/position). Die shot of the CPU as background \cite{die_shot}}
    \label{fig:scan}
\end{figure*}

\subsection{Scan for \ac{EMFI} Susceptible Areas}
\label{sec:experiments:scan}
Multiple parameters like the spatial position of the injection tip, pulse width, pulse voltage, and trigger delay must be determined for a successful attack.
Brute-forcing all of these parameters is impractical because of the huge number of possibilities.
Therefore, we first explored the die surface for \ac{EMFI} susceptible regions using a counter loop that returns a wrong value in case of a successfully injected fault.
This code increases a counter value stored in a register by looping a specified number of cycles while an EM pulse is triggered.
Afterward, the counter is compared to an expected value, and if the expected value and counter differ, we can assume that a fault was injected.
During this experiment, the following procedure is repeated a hundred times at every intersection of a \SI{1}{\milli\metre} grid on the CPU surface:
At first, the injection probe tip is moved to the desired position, and the pulse generator is charged.
Then, the Teensy powers on the target, 
simultaneously waits for SPI messages, and then triggers an EM pulse.
Afterward, the main control unit checks if a fault was injected successfully and powers the target off.

We optimized the EM pulse parameters of the ChipShouter to have the highest success rate. 
Furthermore, we tested several injection tips (\SI{1}{\milli\metre}/\SI{4}{\milli\metre} clockwise/counterclockwise coil) combined with a configured pulse voltage of \SI{500}{\volt} and a pulse width of \SI{40}{\nano\second} for the \SI{1}{\milli\metre} tips and \SI{73}{\nano\second} for the \SI{4}{\milli\metre} tips. 
We moved the injection tip as close as possible to the die to increase the probability of success even further. 
This experiment was performed with each injection probe tip on a \SI{1}{\milli\metre} grid \textbf{without success}. 
No effects on the \ac{DUT} were observed at all.
Thus, we assumed that no fault was injected.
One complete chip scan took about \SI{25}{\hour} to finish.

To increase the success probability further, we lowered V\textsubscript{SoC} from \SI{0.9}{\volt} to \SI{0.59}{\volt} as proposed by Liao et al. \cite{liao_methodology_2019}.
We determined this voltage by lowering V\textsubscript{SoC} until the AMD-SP failed to boot (\SI{0.56}{\volt}), chose a voltage slightly above this threshold and confirmed that the co-processor still boots and works reliably.
The experiment was repeated with all four injection probe tips and the same pulse parameters.
Each experiment showed that \textbf{faults were injected successfully at multiple positions} but the \SI{4}{\milli\metre} clockwise injection probe tip had the most impact. 
\Cref{fig:scan} depicts where on the die and how many faults occurred.

\subsection{Faulting the SRAM}
Having confirmed that the \ac{AMD-SP} is generally vulnerable to \ac{EMFI}, we tried injecting faults into its SRAM to, e.g., flip one or multiple bits.
To detect altered values in the SRAM, we again used the capability to execute our own code: 
With the cache disabled, it copies the same value n times onto the stack and then waits for a specified amount of time before reading back the values.
Meanwhile, an EM pulse is triggered, attempting to inject a fault into the SRAM.
If read values differ from the written ones, a fault counter is increased, and both, the fault counter and values, are printed via UART.
For this experiment, the \SI{1}{\milli\metre} clockwise injection probe was attached to the ChipShouter while pulse parameters remained \SI{500}{\volt} pulse voltage and \SI{40}{\nano\second} pulse width.

\subsection{Faulting the ARK Verification}
\label{sec:experiements:ark}
As explained in \Cref{sub:background:asp}, the on-chip bootloader of the \ac{AMD-SP} loads and verifies the \ac{ARK}.
Thus, it first loads the \acl{ARK} from the connected SPI flash and computes a hash of the key.
This hash is compared to an immutable value stored in an internal ROM. 
If the key is replaced or modified, this check fails, and execution stops.
An adversary who can bypass or alter the hash comparison could also inject his \ac{ARK} to let the \ac{AMD-SP} execute firmware signed by themselves \cite{buhren_one_2021}.

\renewcommand{\arraystretch}{1.3}
\begin{table}[tb]
    \centering
    {\footnotesize
    \begin{tabular}{c|c|c}
        Delay/$\Delta$Delay & Success/Attempts & Success Rate \\
        \hline\hline
        128/4    & 158/10000  & 1.58\%  \\
        \textbf{2364/4} & \textbf{2206/10000} & \textbf{22.06\%} \\
        2384/4   & 352/10000  & 3.52\%  \\
        2391/2   & 68/10000   & 0.68\%  \\
    \end{tabular}
	}
    \caption{Waiting period/window (in wait cycles) and success rates of \ac{EMFI} attack on \ac{ARK} verification}
    \label{tab:ark_attack_success_rates}
\end{table}

We used \SI{500}{\volt} pulse voltage, \SI{73}{\nano\second} pulse width, and the \SI{4}{\milli\metre} clockwise injection probe tip.
The area previously discovered to be most susceptible to \ac{EMFI} (\Cref{sec:experiments:scan}) served as a starting point to improve the position with a finer \SI{0.5}{\milli\metre} grid scan further.
Subsequently, we attempted to fault the \ac{ARK} verification at the refined point that is most susceptible to \ac{EMFI}.
Furthermore, we brute-forced the timing of the EM pulse after the generated SPI trigger, as the correct delay is unknown but limited because the regular verification takes \SI{53}{\micro\second}.
To incorporate all found delays, we chose the median delay of each group of similar delays and a delay window around it ($\Delta$Delay).
Multiple working parameters were found and evaluated for their success rates (\Cref{tab:ark_attack_success_rates}).
The results illustrate that precise timing is required to attack the hash comparison, supposedly because only bypassing or altering some specific instructions in a tiny time frame lead to a successful attack.

\section{Discussion}
\label{sec:discussion}
\subsection{Spatial Resolution and Localization Capabilities}
The spatial resolution of our setup is much higher than needed for the presented attack (\Cref{sec:experiments}).
We scanned the \ac{SOC}'s surface using a \SI{1}{\milli\metre} grid and raised the resolution to a \SI{0.5}{\milli\metre} grid to scan for more susceptible positions around the previously discovered positions to refine the coordinates of the most vulnerable position.
If a high spatial resolution is necessary also depends on the diameter of the injection probe tips.
A large diameter creates a greater EM field that is less location-dependent considering the injected faults.

Scanning the entire chip surface showed that the fault locations are distributed over most parts of the chip (see \Cref{fig:scan}).
Therefore, localization or identification of the \ac{AMD-SP} on a die shot is not possible using only \ac{EMFI}.
Nevertheless, our fault injection experiments revealed that susceptible areas differ based on the targeted component (SRAM vs. register counter loop).

\subsection{Electromagnetic vs. Voltage Fault Injection}
Previous work showed that the \ac{ARK} verification of AMD's Zen CPUs is vulnerable to voltage \ac{FI} \cite{buhren_one_2021}.
We replicated the voltage \ac{FI} attack using our target to compare \ac{EMFI} and voltage \ac{FI}.
The success rate of the voltage \ac{FI} attack is 0.49\% (98 successful glitches in 20000 attempts).
As described in \Cref{sec:experiements:ark}, the success rate of the \ac{EMFI} attack is by a factor of 45 higher (0.49\% voltage \ac{FI} vs. 22.06\% \ac{EMFI}).
While our \ac{EMFI} attack provides a much higher success rate, it also requires a more complex attack setup.
Lowering V\textsubscript{SoC} requires similar hard- and software requirements as a full voltage \ac{FI} attack.
While AMD's mobile CPUs are not equipped with a heat spreader, our target CPU had to be \emph{delidded} for a successful attack, which is not required for voltage \ac{FI}.

\subsection{Lower SoC Voltage vs. Higher Pulse Voltage}
Instead of lowering V\textsubscript{SoC}, one could attempt to use a higher pulse voltage.
The ChipShouter is limited to \SI{500}{\volt}, but other pulse generators support a higher voltage of \SI{750}{\volt} (Avtech pulse generator \cite{Toulemont2020ASP}) or even \SI{1200}{\volt} (SiliconToaster \cite{abdellatif_silicontoaster_2020}).
Recently, researchers discovered that injecting faults into an Intel i3 CPU requires a pulse voltage of \SI{600}{\volt} which is higher than the configured \SI{500}{\volt} of the ChipShouter \cite{trouchkine_2021_intel_i3}.
As the packaging of AMD and Intel CPUs are similar, our attack could benefit from a higher pulse voltage because the hard- and software requirements for lowering the \ac{SOC} voltage could potentially be omitted.

\subsection{Scan and Attack Duration}
One attack cycle consists of three phases: (1) boot of the AMD-SP, (2) one fault injection attempt, and (3) shutdown of the AMD-SP.
Therefore, both the scan and attack duration depend on the irreducible boot and shutdown duration of the AMD-SP.
During a complete chip scan (\Cref{fig:scan}), this cycle repeats a hundred times per position, making the scan slow.
To decrease the overall scan time, a more powerful pulse generator that allows the injection of multiple same-shaped faults in a small interval could be used.
This would increase the success probability per attack cycle.  %

The duration to successfully fault the ARK verification depends on the attack cycle duration and the success rate.
As one attack cycle takes less than four seconds, a successful attempt occurs within a minute, making the attack feasible.
Both the voltage \ac{FI} and \ac{EMFI} success rates could be improved by triggering the fault injection more precisely, e.g., by using an \ac{FPGA} instead of the Teensy.

\renewcommand{\arraystretch}{1.3}
\begin{table}[tb]
    \centering
    {\footnotesize
    \begin{tabular}{l|r}
        Part                      & Price              \\
        \hline\hline
        ChipShouter NAE-CW520-KIT & \EUR{3149.25}      \\
        Teensy 4.0                & \EUR{24.80}        \\
        Raspberry Pi 3B+          & \EUR{38.40}        \\
        BTT SKR Pro v1.2          & \EUR{69.99}        \\
        TMC2209 Motor Drivers     & \EUR{32.53}        \\
        USB Microscope            & \EUR{59}           \\
        Adafruit MLX90640         & \EUR{77.20}        \\
        Power Supply              & \EUR{39.99}        \\
        Standa 8MT175-100 Stages  & $\sim$\EUR{3000}   \\
        Aluminum Profiles         & \EUR{283.72}       \\
        Aluminum Adapter Plates   & $\sim$\EUR{15}     \\
        Angle Bracket             & \EUR{65}           \\
        Wires                     & $\sim$\EUR{50}     \\
        \hline
        Total                     & \EUR{6904.88}      \\
    \end{tabular}
	}
    \caption{Part prices and total costs}
    \label{tab:costs}
\end{table}

\subsection{Setup Costs and Availability}
Our self-built setup consists of commercially available as well as 3D-printed parts.
Two aluminum adapter plates were specially produced based on our design.
However, since these are simple in design, they should be producible in any metal workshop. 
The total price sums up to around €7k, see \Cref{tab:costs}.
Our motion controller based on 3D-printer hardware is customizable and cheap compared to commercially available solutions.
Riscure's EM Probe Station 5 \cite{riscure_probe_station} has nearly the same step size but a smaller travel range of \SI{50}{\milli\metre} and a smaller working area compared to our setup.
Their pulse generator, the EM-Fi Transient Probe \cite{riscure_emfi_transient_probe}, has a maximum coil voltage of \SI{450}{\volt} and a maximum internal current of \SI{64}{\ampere}.
Thus, ChipShouter and Riscure's pulse generator have similar characteristics but vary greatly in their price.
The EM Probe Station 5 and the EM-FI Transient Probe together cost more than €20k, which is more than double the price of our setup.
In case less accuracy is sufficient, cheaper stages could be integrated to nearly halve the costs \cite{badfet}.
If high accuracy stages are beneficial, a self-built pulse generator could be used to lower the costs \cite{abdellatif_silicontoaster_2020}. 
\section{Conclusion}
\label{sec:conclusion}
In this work, we first demonstrated how an \ac{EMFI} setup could be built with a high spatial resolution and that is large enough to target desktop and server systems.
All setup components are commercially available or 3D-printable, and the deployed software is available as open-source software.
Additionally, all CAD files, the controller software, and a parts list is released along this work.
We used the setup to target the \ac{AMD-SP} inside an AMD Ryzen CPU and were able to inject faults to break out of a loop or to alter values stored in SRAM.
Finally, we successfully break the \ac{ARK} verification of the \ac{AMD-SP} as previously carried out via voltage \ac{FI}.
Our \ac{EMFI} attack offers a significantly higher success rate than the voltage \ac{FI} attack and can be readily adapted for future desktop and server targets.

\ifthenelse{\boolean{cameraready}}{
\section*{Acknowledgment}
The work described in this paper has been supported by the Einstein Foundation in the form of an Einstein professorship -- EP-2018-480.
}{}

\bibliographystyle{bib/IEEEtran}
\bibliography{bib/IEEEabrv, references.bib}

\end{document}